\documentclass[11pt,english]{article}
\usepackage[T1]{fontenc}
\usepackage[latin9]{inputenc}
\usepackage{wrapfig}
\usepackage{textcomp}
\usepackage{graphicx}

\DeclareRobustCommand{\greektext}{%
  \fontencoding{LGR}\selectfont\def\encodingdefault{LGR}}
\DeclareRobustCommand{\textgreek}[1]{\leavevmode{\greektext #1}}
\DeclareFontEncoding{LGR}{}{}

\newcommand{\lyxaddress}[1]{
\par {\raggedright #1
\vspace{1.4em}
\noindent\par}
}

\usepackage{moriond}\usepackage{epsfig}
\def\Journal#1#2#3#4{{#1} {\bf #2}, #3 (#4)}

\def\NIMA{{\em Nucl. Instrum. Methods} A}

\def\PLB{{\em Phys. Lett.}  B}
\def\PRL{\em Phys. Rev. Lett.}
\def\PRD{{\em Phys. Rev.} D}

\def\APJ{\em Astrophys. J.}
\def\APP{\em Astropart. Phys.}
\def\EPJC{{\em Eur. Phys. J.} C}

\def\be{\begin{equation}}
\def\ee{\end{equation}}
\def\bea{\begin{eqnarray}}
\def\eea{\end{eqnarray}}

\usepackage{babel}

\begin{document}
\vspace*{4cm}

\title{ICECUBE AS A DISCOVERY OBSERVATORY FOR PHYSICS BEYOND THE STANDARD
MODEL}

\author{K. Helbing, for the IceCube Collaboration~\footnotemark}

\maketitle

\lyxaddress{\begin{center}
\emph{Department of Physics, University of Wuppertal, }\\
\emph{D-42119 Wuppertal, Germany}
\par\end{center}}

\abstracts{Construction of the cubic-kilometer neutrino detector
IceCube at the South Pole has been completed in December~2010. It
forms a lattice of 5160~photomultiplier tubes monitoring a gigaton
of the deep Antarctic ice for particle induced photons. The telescope
is primarily designed to detect neutrinos with energies greater than
100~GeV from astrophysical sources. Beyond this astrophysical motivation
IceCube is also a discovery instrument for the search for physics
beyond the Standard Model. Owing to subfreezing ice temperatures,
the photomultiplier dark noise rates are particularly low which opens
up tantalizing possibilities for particle detection. This includes
the indirect detection of weakly interacting dark matter, direct detection
of SUSY particles, monopoles and extremely-high energy phenomena.}

\section{Introduction}

\footnotetext{Complete author list at http://www.icecube.wisc.edu/collaboration/authorlists/2011/5.html}

\begin{wrapfigure}{O}{0.5\columnwidth}%
\includegraphics[width=0.5\textwidth]{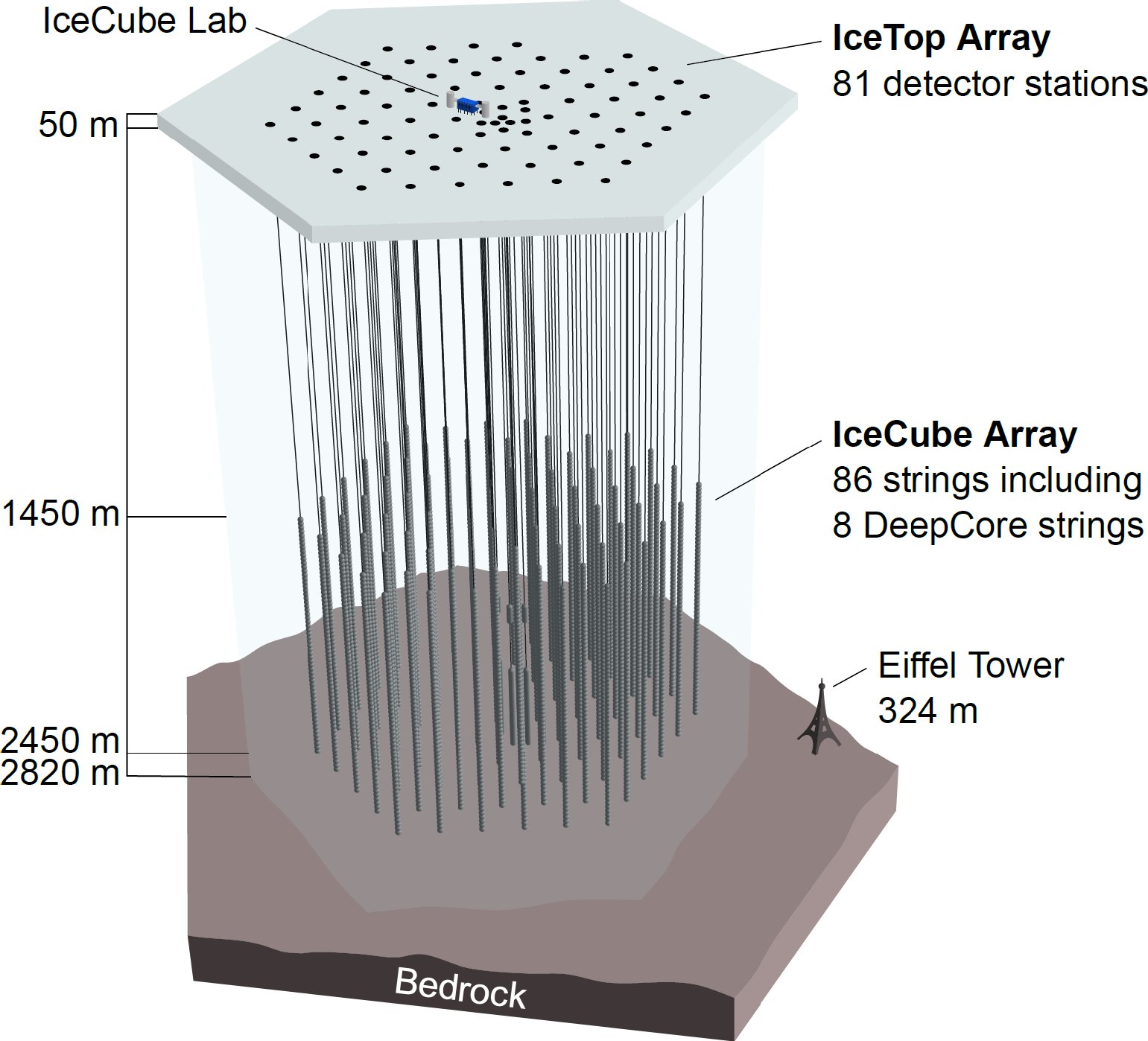}

\caption{The IceCube observatory.\label{fig:IceCube}}
\end{wrapfigure}%
The physics questions that can be addressed with neutrino telescopes
are manifold. They cover the internal mechanisms of cosmic accelerators,
the cosmological evolution of sources, particle physics at center
of mass energies far beyond the TeV scale and the search for new particles
and physics beyond the Standard Model.

\subsection{The detector}

The IceCube Neutrino Observatory at the geographic South Pole has
been completed in December 2010. The detector comprises 5160 digital
optical modules (DOMs) deployed in a three-dimensional array approximately
one cubic-kilometer in size and centered 2 km deep in the clear Antarctic
ice (Fig.~\ref{fig:IceCube}). Each DOM consists of a photo-multiplier
tube and electronics for digitization of waveforms and communication
with neighboring DOMs and the surface. Cherenkov light from the passage
of a relativistic charged particle through the ice creates a pattern
of \textquotedbl{}hit\textquotedbl{} DOMs in the array, and the position
and timing of the hits is used to reconstruct the path of the particle. 

The vast majority of these particles are muons, arriving from cosmic
ray air showers occurring in the atmosphere above the site. IceTop,
the surface component above IceCube, is an air shower array with an
area of 1 km$^{2}$ at a height of 2830~m above sea level. It consists
of 162 ice Cherenkov tanks, grouped in 81 stations. IceTop is primarily
designed to study the mass composition of primary cosmic rays in the
energy range from about 10$^{14}$ eV to 10$^{18}$ eV by exploiting
the correlation between the shower energy measured in IceTop and the
energy deposited by muons in the deep ice.

\section{Astronomy}

\subsection{Neutrino sky}

\begin{figure}
\begin{centering}
\includegraphics[width=0.75\textwidth]{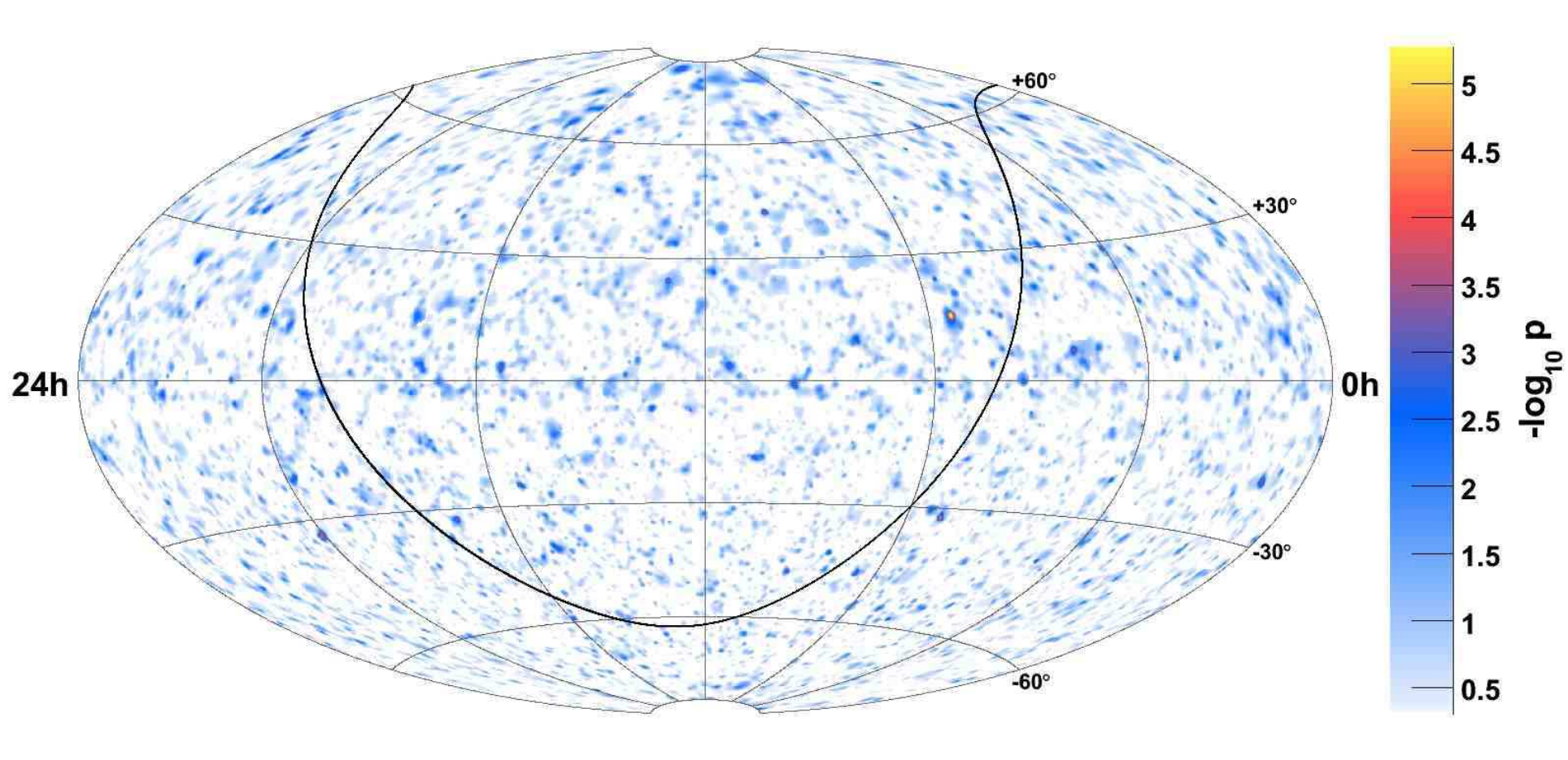}
\par\end{centering}

\caption{Equatorial skymap (J2000) of pre-trial significances (p-value) of
the all-sky point source scan. The galactic plane is shown as the
solid black curve.\label{fig:Equatorial-skymap}}

\end{figure}
IceCube's principal mission is to detect high energy neutrinos from
astrophysical sources. Ultra-high energy cosmic ray (UHECR) experiments
have shown that particles with energies up to a few times $10^{20}$~eV
arrive at Earth. Since the cosmic rays are hadrons also ultra-high
energy (UHE) neutrinos should be produced at these cosmic accelerators.
These neutrinos propagate undeflected through galactic and inter-galactic
magnetic fields and their measurement allows to point back to the
source. Due to the low predicted neutrino fluxes, target masses of
cubic kilometers of water or ice need to be instrumented with photomultiplier
tubes for detection of these neutrinos. 

The detection principle for high energy neutrinos is the measurement
of the Cherenkov light in transparent media which is emitted by charged
leptons produced in neutrino interactions in and around the detector.
The most promising detection channel is muons since muons can propagate
up to several kilometers through the medium. The results of an all-sky
scan~\cite{Abbasi:2010rd} performed with the half-completed IceCube
detector (IC40) are shown in the map of the pre-trial p-values in
Fig.~\ref{fig:Equatorial-skymap}. The most significant deviation
from background is located at 113.75\textdegree{} r.a., 15.15\textdegree{}
dec. The best-fit parameters are 11.0 signal events above background,
with spectral index $\gamma=2.1$. The pre-trial estimated p-value
of the maximum log likelihood ratio at this location is $5.2\cdot10^{-6}$.
In trials using data sets scrambled in right ascension the resulting
post-trial p-value was found to be 18\% -- consequently, the excess
is not claimed. While no TeV neutrinos from astrophysical sources
have been identified yet unambiguously, the partially completed IceCube
detector has set the most stringent upper limits to date.

\subsection{Cosmic rays}

\begin{figure}
\begin{centering}
\includegraphics[width=0.75\textwidth]{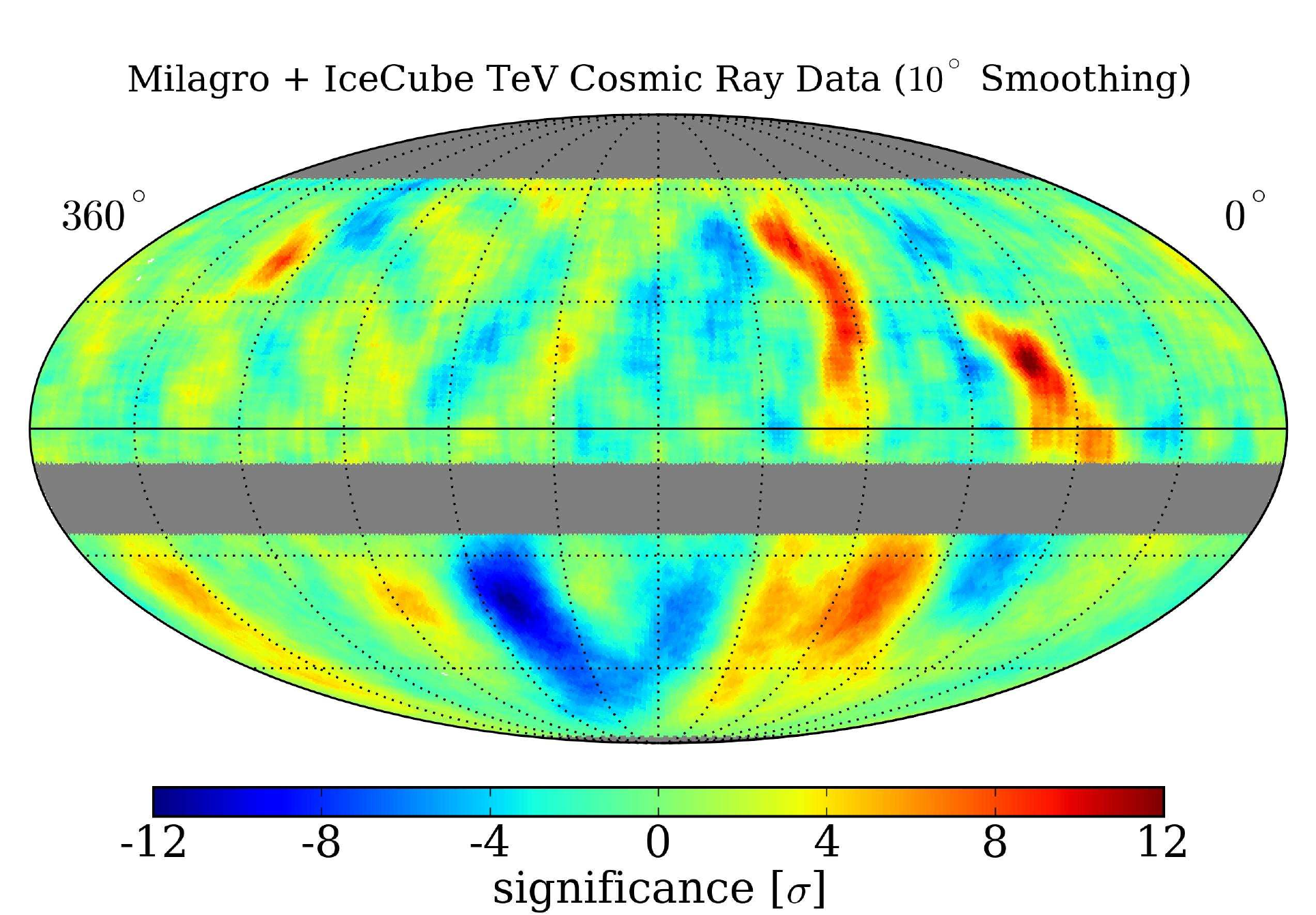}
\par\end{centering}

\caption{Combined map of significances in the cosmic ray arrival direction
distribution observed by Milagro in the northern hemisphere and IceCube
in the southern hemisphere.\label{fig:CRaniso}}

\end{figure}
Between May 2009 and May 2010, the IceCube neutrino detector consisted
of 59 data taking strings recording 32~billion muons. The muons are
generated in air showers produced by cosmic rays with a median energy
of 20~TeV. With this data the southern sky was probed for per-mille
anisotropies in the arrival direction distribution of cosmic rays.
The arrival direction distribution is not isotropic, but shows significant
structure on several angular scales~\cite{CRaniso}. In addition
to a large-scale structure in the form of a strong dipole and quadrupole,
the data show small-scale structures. Fig.~\ref{fig:CRaniso} shows
the combined skymap of significances in the cosmic ray arrival direction
distribution observed by Milagro in the northern hemisphere \cite{Abdo:2008kr}
and IceCube in the southern hemisphere on scales between 15\textdegree{}
and 30\textdegree{}. It exhibits several localized regions of significant
excess and deficit in cosmic ray intensity. The most significant excess
is localized at right ascension 122.4\textdegree{} and declination
\textminus{}47.4\textdegree{} and has a post-trials significance of
5.3\textgreek{sv}. The origin of this anisotropy is unknown.

\section{Searches for non Standard Model particles}

Supersymmetry (SUSY) is currently the most extensively studied amongst
theories beyond the Standard Model (SM). The most direct constraints
on SUSY particle masses have been obtained at LEP and the Tevatron.
While cryogenic dark matter detectors presently have the best sensitivity
for spin independent WIMP-nucleon scattering, indirect searches with
IceCube constrain the spin-dependent cross-sections for neutralino-proton
scattering.

Direct detection channels for SUSY particles are only now being investigated
with the parameter space being largely complementary to that covered
by LHC experiments and WIMP searches -- especially in scenarios where
the gravitino is the lightest SUSY particle. Also, studies of high
light yield exotic signatures from particles like magnetic monopoles
have been performed.\pagebreak{}

\subsection{Indirect WIMP searches}

\begin{wrapfigure}{o}{0.49\columnwidth}%
\includegraphics[width=0.5\textwidth]{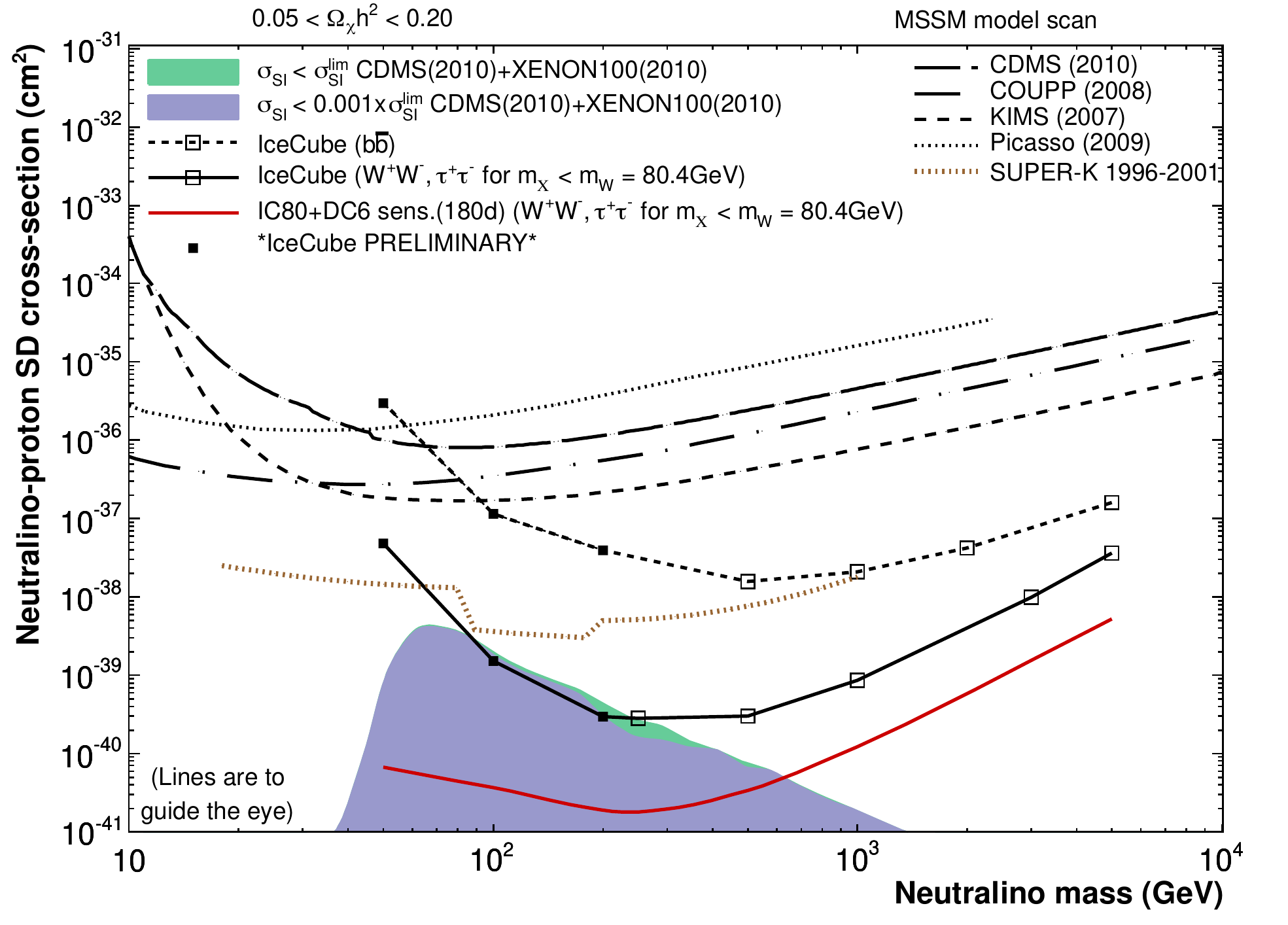}

\caption{Limits on the spin-dependent WIMP-proton cross-section.\label{fig:WimpSun}}
\end{wrapfigure}%

A search for muon neutrinos from neutralino annihilations in the Sun
has been performed with the combined data set of AMANDA and IC22.
No excess over the expected atmospheric background has been observed.
Upper limits have been obtained on the annihilation rate of captured
neutralinos in the Sun and converted to limits on the WIMP-proton
cross-sections. These results are the most stringent limits to date
on neutralino annihilation in the Sun. In Fig.~\ref{fig:WimpSun}
the limits on the spin-dependent WIMP-proton cross-section are compared
with direct search experiments~\cite{Ahmed:2008eu,Behnke:2008,Lee:2007}
and Super-K~\cite{Desai:2004}. Soft WIMP models (annihilation into
$b\bar{b}$) are indicated by the dashed lines, whereas hard models
($W^{+}W^{-}$) are shown in solid lines. Our limits also present
the most stringent limits on the spin-dependent WIMP-proton cross-section
for neutralino masses above 100~GeV. The full IceCube detector with
the densely instrumented DeepCore extension is expected to test viable
MSSM models down to 50 GeV. IceCube is also able to constrain the
dark matter self-annihilation cross section by searching for a neutrino
signal from the Galactic halo~\cite{Abbasi:2011eq}.

\subsection{Direct SUSY searches}

\begin{wrapfigure}{O}{0.49\columnwidth}%
\includegraphics[width=0.49\textwidth,height=0.22\textheight]{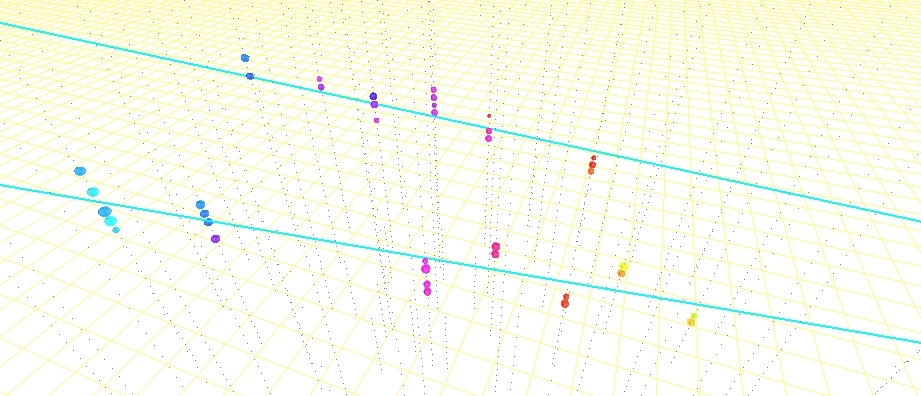}

\caption{Two faint tracks in IceCube from a simulation of parallel staus\label{fig:staus}}
\end{wrapfigure}%
The main phenomenological features of SUSY models arise from the
choice of the symmetry breaking mechanism. Within the minimal supersymmetric
extension of the Standard Model (MSSM) the most extensively studied
mechanisms are gravity mediated supersymmetry breaking and gauge mediated
supersymmetry breaking. In both scenarios the gravitino may be the
lightest supersymmetric particle (LSP). This scenario however, has
not been widely addressed at collider experiments (except in terms
of future concepts) and also WIMP searches usually assume the neutralino
to be the LSP. In that respect a direct search for SUSY with the
gravitino being the LSP is complementary to both ongoing collider
experiments and also to indirect searches.

In models where the lightest supersymmetric particle (LSP) is the
gravitino, typically the next to lightest SUSY particle (NLSP) is
a long lived meta stable slepton (typically a stau). Being charged
the stau is detected by its Cherenkov radiation in the neutrino telescope.
Staus have a small cross section for interactions with {}``normal''
matter. In interactions in the Earth of cosmic neutrinos of typically
PeV energies and above SUSY particles can be produced which eventually
decay into a pair of staus. This pair of staus of a few hundred GeV
mass can propagate through the whole Earth~\cite{Albuquerque:2003mi},
leaving the very distinct signature of two parallel, up-going tracks
separated by several hundred meters when they pass a neutrino telescope
(see Fig.~\ref{fig:staus}). 

This detection signature is quasi background free: Because of the
down-going nature of air shower events, the up-going double stau tracks
are distinguishable e.g. from the high-$p_{T}$ muon events. Upgoing
muon pairs can be created in neutrino-nucleon interactions in the
earth involving charm production and decay~\cite{Albuquerque:2006am}:
$\nu N\to\mu H_{c}\to2\mu\nu_{\mu}H_{x}$. The track length of these
muons is however much shorter than that of staus. Hence their track
separation is smaller as they need to be produced closer to the detector.
Algorithms to identify such stau signatures are currently being developed
for IceCube based e.g. on the track separation and the low brightness.

\subsection{Magnetic monopoles}

\begin{wrapfigure}{o}{0.5\columnwidth}%
\includegraphics[width=0.5\textwidth]{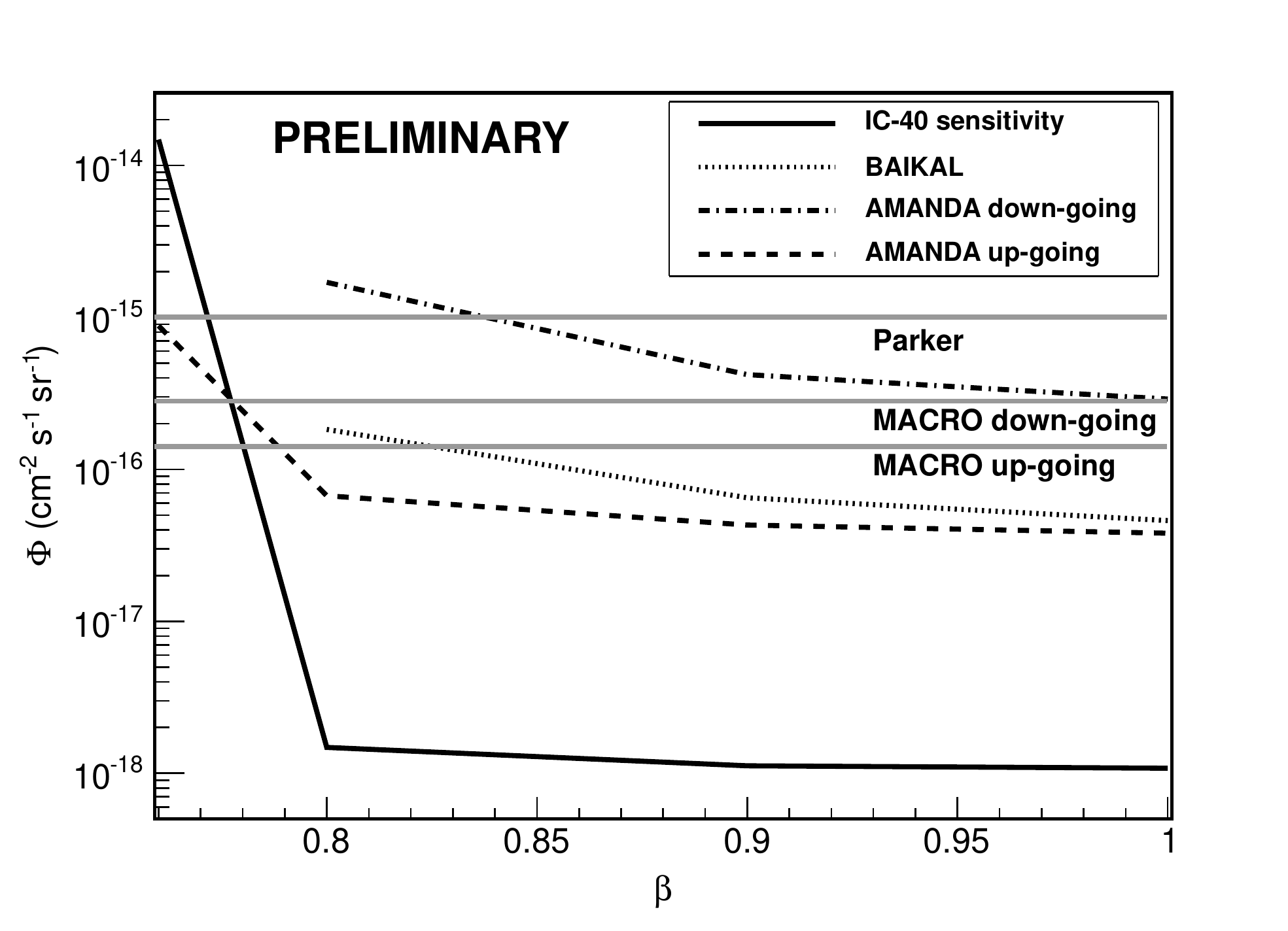}

\caption{Monopole limits and the expected sensitivity of the half completed
IceCube. \label{fig:Monopole-limits}}
\end{wrapfigure}%
Generally, cosmic rays and the big bang are the most likely sources
of massive monopoles, since accelerator energies are likely insufficient
to produce them. The predictions for the mass and charge of monopoles
depend strongly on the choice of the unified group and its symmetry
breaking pattern in the early Universe. The non-observation of the
partner to electric charges may be explained by inflation diluting
the primordial monopole abundance. 

Monopole detectors have predominantly used either induction or ionization
and Cherenkov radiation. Ionization experiments rely on a magnetic
charge producing more ionization than an electrical charge with the
same velocity. The MACRO and Ohya experiments are examples for the
ionization technique~\cite{Ambrosio:2002qq,Orito:1990ny}. 

Large scale Cherenkov telescopes deployed in naturally occurring transparent
media like sea water or glacial ice can detect magnetic monopoles
with both the ionization and Cherenkov radiation from magnetic monopoles:
For relativistic monopoles moving at a speed above the Cherenkov threshold
the light yield is excessive (several thousand times more) compared
to Standard Model particles. But even at velocities below the Cherenkov
threshold monopoles are observable through delta rays and ionization,
again exceeding the light yield of other particles of the same velocity.
Moreover, some GUT theories predict that monopoles catalyze the decay
of nucleons which would be observed by a series of light bursts produced
along the monopole trajectory.

Searches for relativistic monopoles with Cherenkov neutrino telescopes
have already been performed with the AMANDA and BAIKAL detector and
are being investigated with the IceCube detector \cite{Antipin:2007zz,Abbasi:2010zz}.
Fig.~\ref{fig:Monopole-limits} shows that sensitivities well below
the so called Parker bound~\cite{Parker:1970,Turner:1982ag} have
been reached for relativistic monopoles. Parker pointed out that the
abundance of magnetic monopoles cannot be as high as to deplete galactic
magnetic fields. Strategies to identify non-relativistic monopoles
in IceCube are currently being developed. In conclusion, IceCube is
entering the interesting region of sensitivities for monopole searches
spanning a wide range of relativistic and sub-relativistic velocities.

\section{Extremely-high energy neutrinos}

\begin{wrapfigure}{O}{0.5\columnwidth}%
\includegraphics[width=0.5\textwidth]{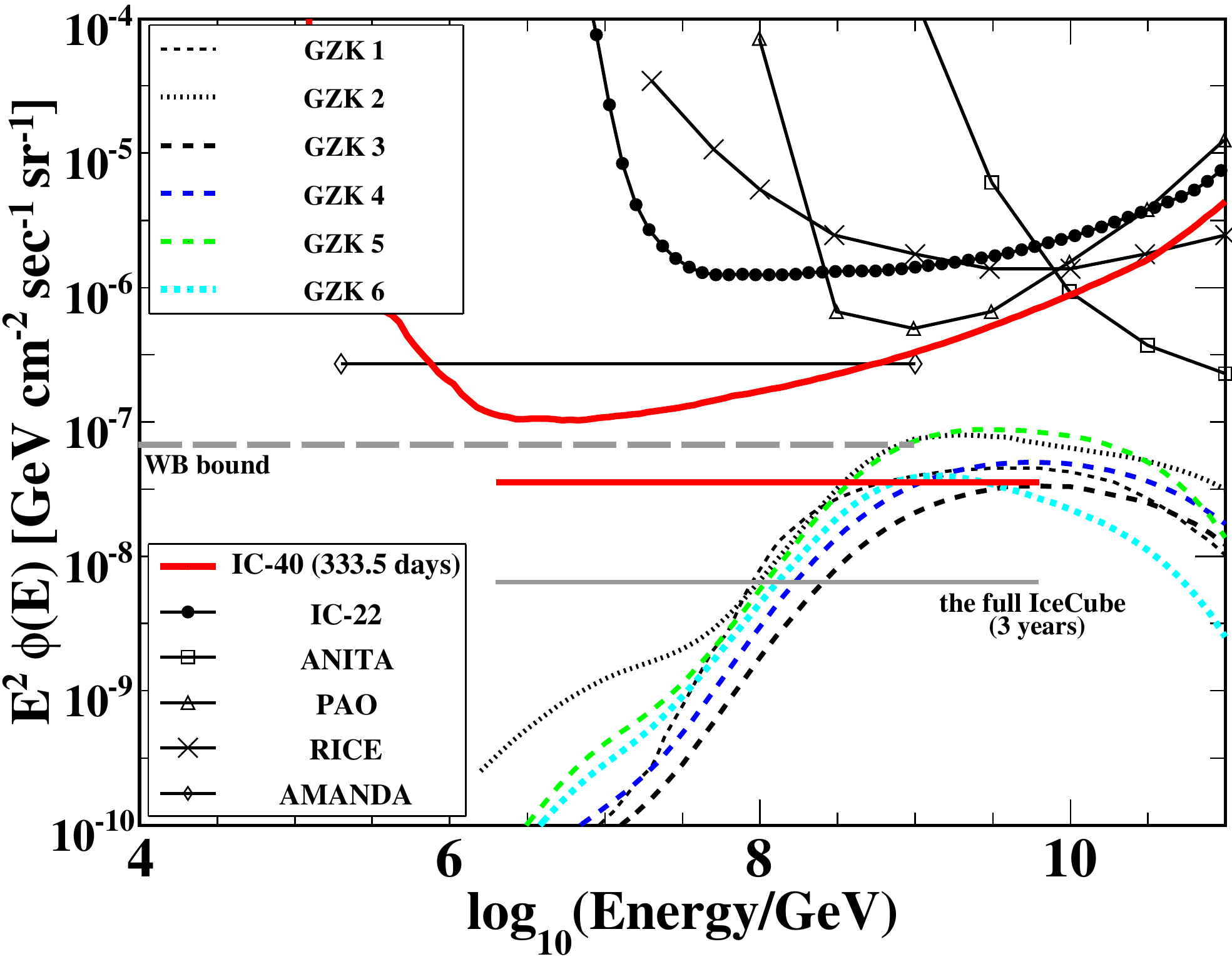}

\caption{Quasi-differential model-independent 90\% CL limit normalized by energy
decade and E$^{-2}$ spectrum integrated limit on all flavor neutrino
fluxes from the 2008-2009 IceCube EHE analysis (red solid lines).
The systematic errors are included. Various model predictions (assuming
primary protons) are shown for comparison.\label{fig:EHE}}

\end{wrapfigure}%
Cosmogenic neutrinos may give a unique picture of the Universe in
the highest energy regime. With the Greisen-Zatsepin-Kuzmin (GZK)
process the highest energy cosmic-rays interact with the cosmic microwave
background producing these neutrinos~\cite{Greisen:1966,Berezinsky:1969}.
Hence, cosmogenic neutrinos carry information about the sources of
the highest energy cosmic-rays, such as their location, cosmological
evolution, and cosmic-ray spectra at the sources. 

On the other hand, tiny departures from Lorentz invariance have effects
that increase rapidly with energy and can kinematically prevent cosmic-ray
nucleons from undergoing inelastic collisions with CMB photons. With
charged cosmic-rays alone it is impossible to differentiate between
a true GZK cutoff or the fading spectrum of cosmological accelerators. 

Underground neutrino telescopes, such as IceCube, can detect EHE neutrino
interactions through the strong Cherenkov radiation emitted by the
charged secondary particles. In a neutrino telescope, an EHE neutrino
interaction is identified by the extremely high number of Cherenkov
photons deposited in the detector. Fig.~\ref{fig:EHE} shows the
search for neutrinos with energies above $10^{15}$~eV using data
collected with the half-completed IceCube detector in 2008\textminus{}2009~\cite{Abbasi:2011ji}.
Our limits are competitive up $10^{19}$~eV.

\subsection{Extensions of IceCube}

Besides the GZK process, neutrinos at ultra-high energies are also
a valuable tool to study the neutrino-nucleon cross section at high
center of mass energies. For energies above $10^{16}$~eV the Standard
Model cross section rises roughly with a power law $\sigma_{SM}\propto E_{\nu}^{0.36}$
in the energy of the neutrino~\cite{Gandhi:1998ri}. Naively, the
cross section for black hole creation scales with the Schwarzschild
radius $\sigma_{BH}\propto r_{S}^{2}\propto E_{cm}^{2}\propto E_{\nu}$
eventually exceeding the Standard Model processes. For a more refined
discussions also addressing extra dimensions see for example~\cite{Anchordoqui:2003jr}.

The detection of the small neutrino flux predicted at the highest
energies (E > 10$^{17}$ eV) requires detector target masses of the
order of 100 gigatons, corresponding to 100 km$^{3}$ of water or
ice. The optical Cherenkov neutrino detection technique is not easily
scalable from the 1~km$^{3}$-scale telescopes to such large volumes.
Several techniques have been studied to realize such huge detection
volumes. Radio Cherenkov neutrino detectors search for radio Askaryan
pulses in a dielectric medium as the EHE neutrino signature~\cite{Askaryan:1962}.
Acoustic detection is based on the thermo-acoustic sound emission
from a particle cascade depositing its energy in a very localized
volume causing a sudden expansion that propagates as a shock wave
perpendicular to the cascade~\cite{Askaryan:1979}.

Within IceCube the properties of the South Pole ice for acoustic~\cite{Abbasi:2009si,Abbasi:2011zy,Abbasi:2010vt}
and radio~\cite{Landsman:2010} detection have been studied with
respect to signal attenuation, refraction and the noise environment~.
The results turn out to be very favorable promising longer signal
attenuation lengths than for the optical detection, allowing for a
sparse instrumentation of the Antarctic ice. Consequently, the installation
of a 80 km$^{2}$ radio array dubbed ARA has commenced~\cite{Allison:2011wk}.
Studies to augment the radio detection with acoustic sensors show
that it may be possible to bootstrap detection strategies for the
large effective volumes by building a hybrid detector~\cite{Besson:2009}.
A signal seen in coincidence between any two of the three methods
(radio, acoustic, optical) would be unequivocal. The information from
multiple methods can also be combined for hybrid reconstruction, yielding
improved angular and energy resolution.

Another addition pursued is the RASTA detector which will complement
the IceTop air-shower detector with an extended surface array of radio
antennas~\cite{Boser:2010sw}. Besides the additional capabilities
for cosmic-ray composition studies, this combination also enhances
IceCube's optical high-energy neutrino sensitivity by vetoing the
air-shower background.

\section*{Acknowledgments}

KH acknowledges the support from German Ministry for Education and
Research (BMBF).

\end{document}